\documentstyle[referee]{mn}

\footnotesize
\newdimen\minuswidth    
\setbox0=\hbox{$-$}
\minuswidth=\wd0
\catcode`@=\active
\def@{\kern\minuswidth}
\newdimen\digitwidth    
\setbox0=\hbox{\rm0}
\digitwidth=\wd0
\catcode`!=\active
\def!{\kern\digitwidth}
\normalsize

\title[Determination of the orbital parameters of binary
pulsars]{Determination of the orbital parameters of
binary pulsars}
\author[Freire, Kramer and Lyne]{
P. C. Freire,
M. Kramer,
A. G. Lyne\\
University of Manchester, Jodrell Bank Observatory, Macclesfield,
Cheshire, SK11~9DL, UK
}

\begin{document}

\maketitle
\newcommand{\setthebls}{
}

\setthebls

\begin{abstract}
We present a simple novel method for determining the orbital
parameters of binary pulsars. This method works with any sort of orbital
sampling, no matter how sparse, provided that information on the period
derivatives is available with each measurement of the rotational period of
the pulsar, and it is applicable to binary systems with nearly circular
orbits. We use the technique to precisely estimate the hitherto unknown orbital
parameters of two binary millisecond pulsars in the globular cluster
47~Tucanae, 47~Tuc~S and T. The method can also be used more generally
to make first-order estimates of the orbital parameters of binary
systems using a minimal amount of data.
\end{abstract}


\section{Introduction }
\label{sec:introduction}

The discovery of a pulsar whose period changes significantly usually
indicates that it is a member of a binary system.  It is then
important to determine the Keplerian orbital parameters of the system
in order to investigate the astrophysics of the two stars and to
obtain a coherent timing solution for the rotation of the pulsar.  The
latter provides precise astrometric information as well as the period
derivative which may give information on the magnetic flux density and age
of the pulsar or, if it is in a globular cluster, on the gravitational
field of the cluster \cite{and92,phi92b}.

The usual procedure for obtaining orbital parameters involves fitting
a Keplerian model to a series of period measurements.  Such a
procedure works well, provided that it is possible to determine the
rotational period on several occasions during a single orbit.
However, there are often circumstances where this is not the case,
such as where interstellar scintillation permits only sparse positive
detections of a pulsar \cite{clf+00}.

We have therefore sought a way of employing the hitherto unused information
contained in the measured orbital accelerations to obtain a full, coherent
solution for the Keplerian orbital parameters.

In what follows, we present a simple procedure for estimating the
orbital parameters of a binary that is completely independent of the
distribution of the epochs of the individual observations, provided
that the period derivatives are known in each observation. In section
\ref{sec:general}, we present the equations for the period and
acceleration of a pulsar in an eccentric binary system as a function
of its position in the orbit. We also provide an analysis of the
circular case and show how the orbital parameters can be obtained from
a plot of the acceleration of the pulsars against rotational period.
In section \ref{sec:app}, we demonstrate the application of the
procedure to a number of pulsars in the globular cluster 47~Tucanae
which have known orbital parameters, and then to 47~Tuc~S and T,
neither of which hitherto had an orbital solution.  In section
\ref{sec:refining}, we show how to improve the estimates of the
orbital parameters, and proceed to refine the orbits of 47~Tuc~S and T.

\section{Binary orbits in the acceleration/period plane}
\label{sec:general}

\begin{figure*}
\setlength{\unitlength}{1in}
\begin{picture}(0,5)
\put(-2.5,0){\includegraphics{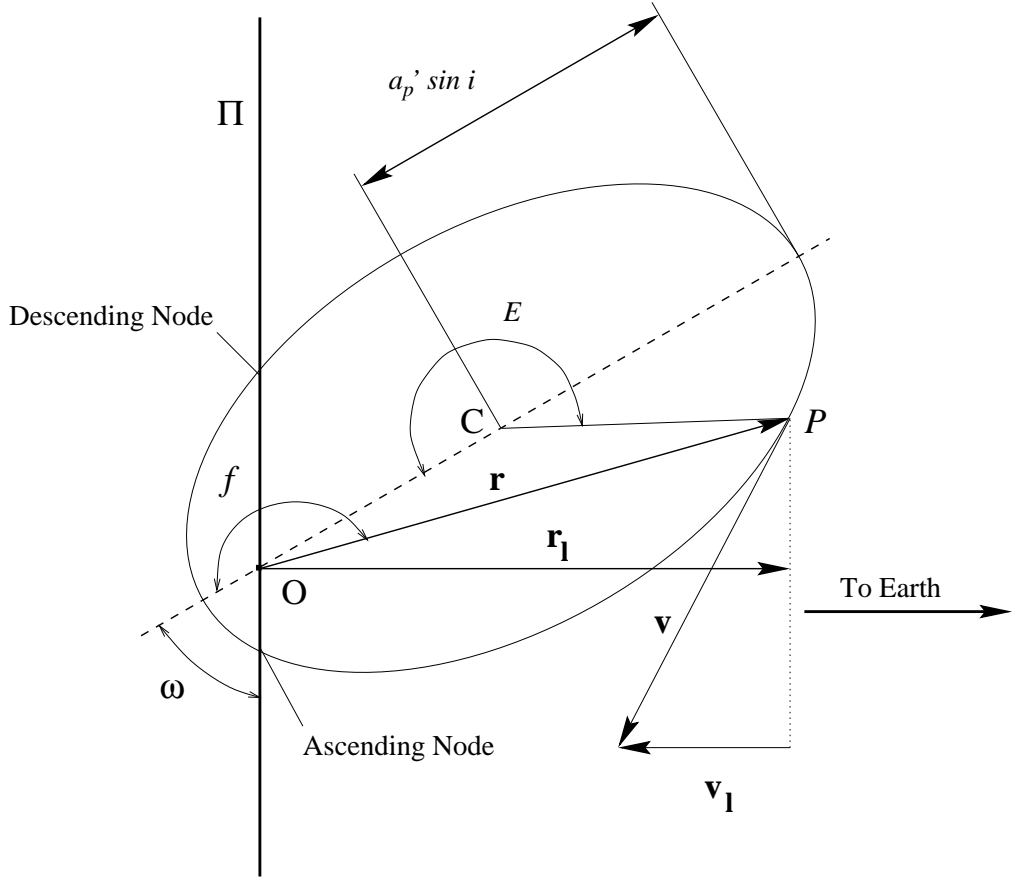}}
\end{picture}
\caption
[]
{Geometrical parameters for an elliptical orbit.}
\label{fig:orbit}
\end{figure*}

Figure \ref{fig:orbit} represents the orbit of a binary pulsar (point P)
around the centre of mass of the system (O) projected onto a plane that
contains the direction towards the Earth (i.e., perpendicular
to the plane of the sky, $\Pi$), and the line of nodes where the
orbital plane intersects $\Pi$. The orbit is an ellipse of
eccentricity $e$ and projected semi-major axis $a_P = a_P' \sin i $
($i$ is the unknown inclination of the orbit relative to the plane of the
sky). {\bf r} is the vector connecting O to P, and $r\, = \, | {\bf r} |$
is the distance of the pulsar to O, $r_{l}\, = \, | {\bf r_l} |$ is the
projection of this to the line-of-sight. $\omega$ is the longitude of
periastron and $f$ is the angle of the pulsar to the periastron measured
at O, also called "true anomaly". According to Roy (1988), the
equation for $r_{l}$ as a function of $f$ is given by

\begin{equation}
r_{l}(f) = r \sin ( \omega + f) =
a_{P}  \frac{ \sin ( \omega + f )}{1 + e \cos f}
\label{eq:rl}
\end{equation}

The time derivative of $r_{l}$ is the line-of-sight (or radial) velocity,
$v_{l}$. Using the results in Roy (1988)\nocite{roy88}, we obtain:

\begin{equation}
v_{l}(f) = \frac{2 \pi}{P_{B}} \frac {a_{P} } {\sqrt  {1 - e^2}  }
\left\{  \cos (\omega + f) + e \cos \omega \right\}
\label{eq:vl}
\end{equation}
where $P_B$ is the orbital period of the binary.
The apparent rotational period of the pulsar $P$ as a function of $f$
and the intrinsic rotational period $P_0$ is given by
\begin{equation}
P(f) = P_{0} \left( 1 + \frac{v_l(f)}{c} \right) \left( 1 - \frac{v^2(f)}{c^2}\right)^{-1/2}
\simeq P_{0} \left( 1 + \frac {v_ {l}(f)}{c} \right)
\label{eq:P}
\end{equation}
if the total velocity of the pulsar, $v(f)$ is small compared to $c$.

Differentiating equation \ref{eq:vl} in time, we obtain the acceleration
along the line of sight $A_{l}$ as a function of $f$
\begin{equation}
A_{l}(f) = -\left( \frac{2 \pi}{P_{B}} \right)^2 \frac{a_{P} }{1 - e^2}
(1 + e \cos f)^2 \sin(\omega + f)
\label{eq:Accel}
\end{equation}

\begin{figure*}
\setlength{\unitlength}{1in}
\begin{picture}(0,5.5)
\put(-3.7,-0.3){\includegraphics{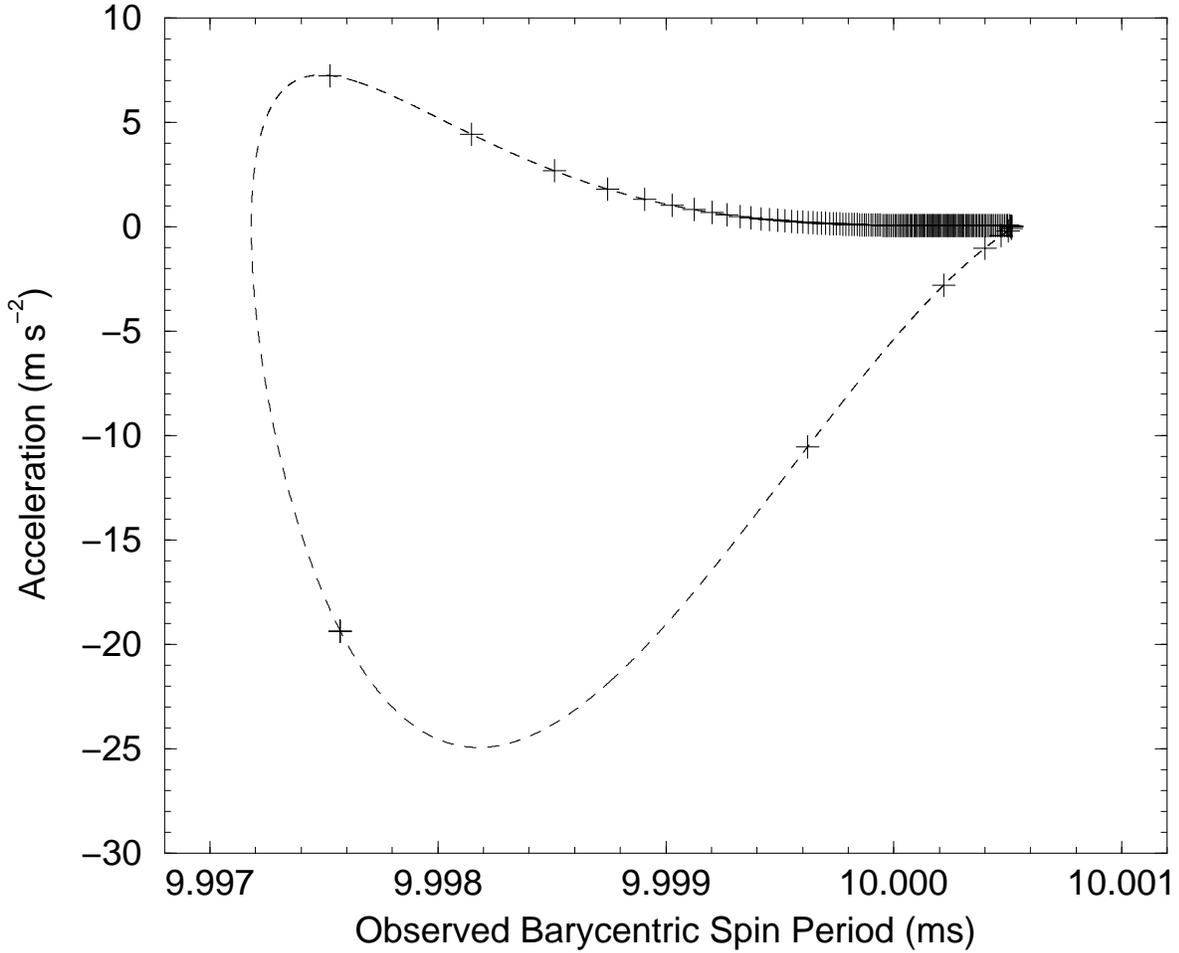}}
\end{picture}
\caption
[]
{A simulated binary system with an orbital period of 1 day, $x$ = 1s,
$e$ = 0.9 and $\omega \, = \, 140^\circ$. The pulsar has an intrinsic
rotational period of 10 ms. The dashed line is the acceleration as a
function of rotational period measured at the barycentre of the solar
system. This pulsar was ``observed'' every ten minutes, each
observation is represented as a $+$ sign. The zones with a
smaller concentration of observations correspond to the periastron, where the
pulsar spends relatively little time, but where accelerations are
largest. The cusp is slightly to the right of apastron, which is in
the near straight segment where the acceleration is close to zero
and where the density of observations is greatest.}
\label{fig:accp}
\end{figure*}

Plotting $A_{l}(f)$ as a function of $P(f)$, we obtain a parametric curve
that does not depend on time, and therefore does not require the solution
of Kepler's equation. This curve is illustrated in Figure
\ref{fig:accp} for a binary pulsar of spin period 10 ms and orbital 
parameters $e$ = 0.9, $\omega = 140 ^ \circ $,
$P_{B}$ = 1 day and $x\, \equiv\, a_P/c =  a_{P}'\, \sin i  / c = 1 s$.

All known binaries containing "true" millisecond pulsars, i.e., those with
periods below 20 ms,
have rather circular orbits \cite{cam95,lyn95}. For these, we can set
$e = \omega$ = 0 in the equations above, so that
equations \ref{eq:P} and \ref{eq:Accel} reduce to

\begin{equation}
P(f) = P_{0} + P_{0} x \frac{2 \pi} {P_{B}} \cos f  \equiv P_{0} + P_{1} \cos f
\label{eq:Pcirc}
\end{equation}
and
\begin{equation}
A(f) = - \frac {4 \pi^2}{P_{B}^2} x c  \sin f \equiv -A_{1} \sin f.
\label{eq:Acirc}
\end{equation}

The track followed by such pulsars in the period/acceleration space is
thus an ellipse centred on the point ($P_ {0}$, 0) and having as
horizontal and vertical semi-axes the values $P_{1}$ and $A_{1}$
respectively, with the pulsar moving in a clockwise direction.

Generally, once an ellipse has been found to fit a set of measurements
of barycentric period and acceleration for a given pulsar, we can
easily recover the two relevant orbital parameters in a circular orbit:

\begin{equation}
P_B =  \frac {P_1}{P_0} \frac {2 \pi c}{A_1}
\label{eq:PB1}
\end{equation}

\begin{equation}
x = \left(  \frac{P_1}{P_0} \right)^2 \frac {c}{A_1}
\label{eq:X1}
\end{equation}

Using these newly determined orbital parameters, we can calculate the 
angular orbital phase for each $k$th data point, i.e.~for each pair of
acceleration and period measured ($A_k$, $P_k$):

\begin{equation}
\phi_k = \arctan \left( - \frac {A_k}{A_1}   \frac{P_1}{P_k - P_0}  \right)
\label{eq:phase}
\end{equation}

Since the time $T_k$ of each observation is also known, we can determine the
time of the nearest ascending node, or simply $T_{\rm asc, k}$ for each
observation:

\begin{equation}
T_{\rm asc, k} = T_k - \frac{\phi_k}{2 \pi} P_B
\label{eq:T0}
\end{equation}
We can use these values to refine the orbital period
(section \ref{sec:refining}).

\section{Application of the Procedure to Pulsars in 47~Tucanae}
\label {sec:app}

In this section, we first test the procedure described above on a
number of known binary systems in the globular cluster 47~Tucanae and
then apply it to two systems with hitherto unknown orbits.

Between August 1997 and August 1999, the 64-m Parkes radio telescope
was used to make a total of 108 observations of the pulsars in the
globular cluster 47~Tucanae (henceforth 47~Tuc) at a central frequency
of 1374\,MHz ($\lambda = 21$\,cm).  About half of these observations
were of 4.6 hours duration, the remainder were shorter.
These data have been used to make accurate timing measurements of
the pulsars in 47~Tuc, the long integrations being required in order
to detect the very faint pulsars (Freire et al. 2000)\nocite{fcl+00b}.
However, it can also be used to search for previously undetected pulsars
\cite{clf+00} using an analysis which searches over a range of
different accelerations.

In a first search, targeted at binaries with periods as short as 90
minutes, each of these 4.6-hour observations was divided into sixteen
17.4-minute segments, and an acceleration search was performed on each
segment. The trial accelerations ranged from $-$30 m s$^{-2}$ to 30 m
s$^{-2}$, with a step of 0.3 m s$^{-2}$. This search was remarkably
successful, with the discovery of nine new millisecond pulsars,
all of which are binary \cite{clf+00}. This increased the
total number of known pulsars in 47~Tuc to 20. All have periods between
2 and 8 ms, and 13 are members of binary systems.

The reason that new pulsars are still being found in 47~Tuc is
that the pulsars scintillate strongly at 1374 MHz, with
flux density variations over more than one order of magnitude in a few
hours. This occasionally increases the flux density of a weak pulsar
above the sensitivity threshold of the Parkes radio telescope, making
it detectable for perhaps one or two hours in the 17.4-minute
sub-integrations.

If a newly discovered binary system has an orbital period comparable
to the length of an observation, then it is possible to make a unique
determination of the orbital parameters of the pulsar using data from
a single observation.  However, some of the newly-discovered binaries
(e.g. 47~Tuc~S and T) have longer orbital periods. Thus far, it has
been impossible to determine the orbital parameters of these binaries,
because we cannot choose when to observe them - that is primarily
determined by scintillation, and detections of these pulsars are very
rare (47~Tuc~S was detected on a single observation, and 47~Tuc~T on
three observations in the 17.4-minute search). More importantly, the
extremely sparse successful detections typically occur once every few
hundred orbits.  For these two pulsars, the number of detections has
been increased to 4 and 7 respectively, by dividing the observations
into 4 70-minute segments and conducting acceleration searches on each
of these, so increasing the sensitivity by a factor of 2 over the
search of Camilo et al. (2000).  These detections are summarised in Table
\ref{tab:detections}, but unfortunately are still much too sparse to
determine a unique orbit from the period determinations alone.

\begin{table}
\begin{center}
\begin{tabular}{ c c c c r }
\hline
Pulsar & $T$   & $P$  & A            & (S/N) \\
       & (MJD) & (ms) & (m s$^{-2}$) &     \\
\hline
\hline
47 Tuc S
  & 50741.644 & 2.8304388 & $-0.78\pm 0.06$ & 16 \\
  & 50741.693 & 2.8304060 & $-0.86\pm 0.06$ & 14 \\
  & 51000.765 & 2.8304720 & @$0.72\pm 0.04 $ & 14 \\
  & 51000.814 & 2.8304981 & @$0.58\pm 0.04 $ & 11 \\
  & 51002.899 & 2.8303046 & @$0.56\pm 0.08 $ & 9  \\
  & 51216.176 & 2.8305188 & $-0.50\pm 0.08 $ & 9 \\
\hline
47 Tuc T
  & 50740.654 & 7.5878889 & $-0.92\pm0.08 $ & 12 \\
  & 50740.703 & 7.5878334 & $-0.24\pm 0.08$ & 13 \\
  & 50746.626 & 7.5884100 & @$1.68\pm 0.04$  & 12 \\
  & 50746.675 & 7.5885810 & @$1.67\pm 0.06$  & 19 \\
  & 50981.811 & 7.5878836 & @$0.72\pm 0.06$  &  9 \\
  & 50983.932 & 7.5878552 & $-0.44\pm 0.08$ & 11 \\
  & 51005.893 & 7.5891029 & @$0.52\pm 0.08$  & 13 \\
  & 51040.857 & 7.5891350 & @$0.00\pm 0.04$  & 66 \\
  & 51335.904 & 7.5891358 & @$0.00\pm 0.02$  & 11 \\
\hline
\end{tabular}
\caption{ Measured periods and accelerations for 47~Tuc~S and T. S is
detected on four different observations of 47~Tuc,
T is detected on 7 different observations. This represents a total of
6 and 9 measurements of period and acceleration respectively. The
signal-to-noise ratio (S/N) varies due to scintillation.}

\label{tab:detections}
\end{center}
\end{table}

\begin{figure*}
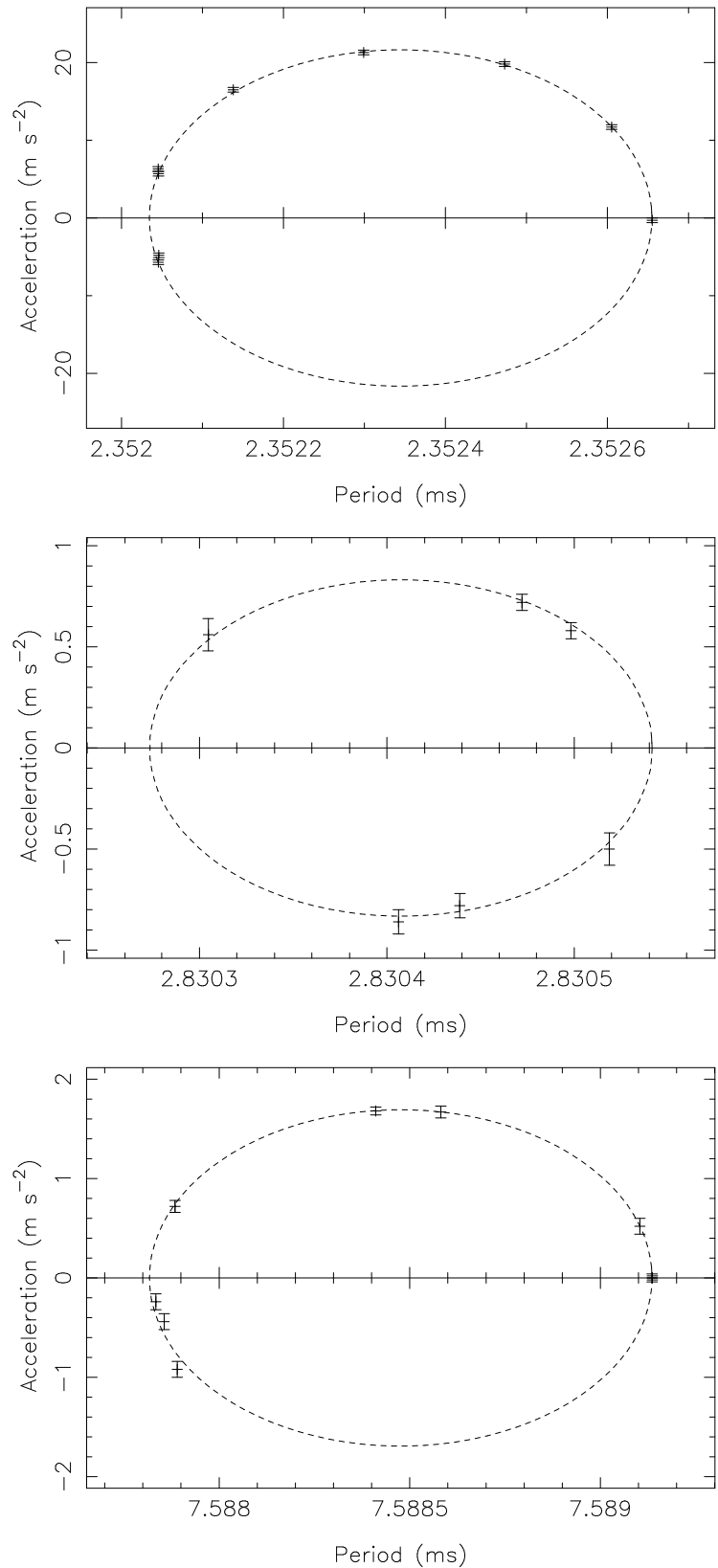

\setlength{\unitlength}{1in}
\begin{picture}(0,8.7)
\put(-2.5,8.9){\includegraphics{W.ps}}
\put(-2.5,6.0){\includegraphics{S.ps}}
\put(-2.5,3.1){\includegraphics{T.ps}}
\end{picture}
\caption
[] {Observed accelerations plotted against barycentric spin period for
47~Tuc~W (top), 47~Tuc~S (middle) and 47~Tuc~T (bottom), three binary
millisecond pulsars in the globular cluster 47~Tuc. The latter two
plots were obtained from the numerical data in Table
\ref{tab:detections} and the dashed ellipses represent the best fits.}
\label{fig:acc_period}
\end{figure*}

The top diagram in Fig. \ref{fig:acc_period} shows a plot of the
measured accelerations as a function of barycentric period for
47~Tuc~W, a 2.35-ms pulsar in a 3.2-hour binary system. The dashed
ellipse represents the best fit to the data obtained using the
implementation described in Appendix~A.  The fit clearly describes the
data well.  For this particular pulsar, we do not observe large
negative accelerations which would correspond to times when the pulsar
lies behind the companion with respect to Earth.  This apparent lack
of negative accelerations is due to the known eclipses in this system,
which cover about 40 \% of the whole orbit.

As well as 47~Tuc~W, we have repeated the analysis on 47~Tuc~J and
47~Tuc~U and summarise the results from the fitting procedure in Table
\ref{tab:parameters}.  Given in the table are the fitted parameters
which can be compared with those obtained form the full, coherent
solutions which are of course several orders of magnitude more
precise. The fitted values are all consistent with these within the
quoted errors.

The other two diagrams in Fig. \ref{fig:acc_period} show the data
presented in Table \ref{tab:detections}.  These few and very sparse
detections clearly delineate well-defined ellipses and have proved
enough to make preliminary determinations of the orbits of both of
these pulsars, which we made using the method outlined in section
\ref{sec:general} and Appendix A.  The orbital parameters resulting
from the best fitting ellipses to the data from these two pulsars are
given in Table \ref{tab:parameters}.

\section{Refining the ephemeris}
\label{sec:refining}

As mentioned at the end of section \ref{sec:general}, it is possible
to refine the orbital period further using the local values of the
$T_{\rm asc,k}$, in the same manner that a set of times-of-arrival of
pulses from a pulsar (TOAs) can be used to improve the estimate of the
rotational period of an isolated pulsar, once a reasonable estimate of
the rotational period is available. This also allows the determination
of a more precise value for a global $T_{\rm asc}$.

As in the pulsar case, the problem is that we don't know precisely
how many orbits have occurred between any two observations.
With data as sparse as that of 47~Tuc~S and T, this task is normally
impossible. However, we already know approximate orbital periods for
the binary systems from the procedure outlined in section
\ref{sec:general} and implemented in Appendix A. We also know more than
two times of ascending node for each binary, calculated using
equation \ref{eq:T0}.  Thus, all we have to do is to estimate
a binary period and global $T_{\rm asc}$ that produce the observed
$T_{\rm asc,k}$, with the restriction that the orbital period
should be consistent with that determined in the previous section.

We have developed two independent methods to perform this task.
In the first method, we essentially treat each $T_{\rm asc, k}$ 
derived from Eqn.~(\ref{eq:T0}) as a TOA.
For each observation, $k$, we compute a ``residual''
\begin{equation}
R_k = \frac{T_{\rm asc, k} - T_{\rm asc, 1}}{P_B} - 
      \lfloor \frac{T_{\rm asc, k} - T_{\rm asc, 1}}{P_B} \rfloor
\label{eq:residual}
\end{equation}
(where $\lfloor a \rfloor$ is the largest integer smaller than $a$)
and plot it against each $T_{\rm asc, k}$. If the previous estimate of
$P_B$ used in eq. \ref{eq:residual} is correct, then all $R_k$ should
be randomly distributed around zero within the errors. An error in
$P_B$ results in a slope in the data points. Hence, we model this by
performing a least-squares fit of a straight line and repeat this
procedure for a set of trial $P_B$s.  We search a range of $P_B$s
centred on the previous best estimate, such that the total phase
difference over the whole time span, $T_{\rm asc, n} - T_{\rm asc, 1}$ does
not exceed unity (as we can assume that our previous guess was
sufficiently close to the correct value). We note that for each trial
$P_B$, we re-calculate the $T_{\rm asc, k}$'s using the phases, $\phi_k$,
from Eqns.~(\ref{eq:phase}) and (\ref{eq:T0}). The results of the
straight line fit are then used to correct the previous estimates for
$T_{\rm asc}$ and $P_B$.  If the fitted $P_B$ and $T_{\rm asc}$ are at some
point coincident with those of the binary, we should expect to see a
sharp decrease in the residuals given by eq. \ref{eq:residual}.

In the second method, we make direct use of the phase information, $\phi_k$,
by fitting the function
\begin{equation}
\phi = \frac{T-T_{\rm asc}}{P_B} - 
      \lfloor \frac{T-T_{\rm asc}}{P_B} \rfloor
\label{eq:resfunc}
\end{equation}
to our data set of observing epochs and derived phase information
($T_k, \phi_k$). We perform a least squares minimization, where due to
the non-linear nature of Eqn.~\ref{eq:resfunc} we employ a robust
Downhill-Simplex algorithm \cite{ptvf92} to obtain $T_{\rm asc}$ and
$P_B$. We start the simplex routines for various starting points using
$T_{\rm asc, 1}$ and a variation of $P_B$ as in the first method.  In
order to obtain reliable error estimates, we again use a Monte-Carlo
simulation as described in Appendix A, generating a synthetic data set from
the real $(T_k, \phi_k$)'s and the error estimates for the phase as
derived from the first Monte-Carlo application.

Our experience shows that both methods yield very similar results, usually
agreeing within the errors. However, depending on the spread of the data
points, length of total time covered and uncertainties, one or the other
method may yield superior results. While the orbital solution for 47~Tuc~T
was refined using the first method, the second method provided better
results in the case of 47~Tuc~S. Applying both methods at the same time
serves in fact as a good indicator whether the obtained solution is
unique.

\begin{table*}
\begin{center}
\begin{tabular}{ l l l l l l }
\hline
Parameter & J & U & W & S & T \\
\hline
\hline
 & & & & \\
$P_0^a$ (ms) & 2.1006335(6) & 4.342831(1) & 2.3523445(8) & 2.830407(8) &  7.588476(4) \\
$P_0^b$ (ms) & 2.1006336    & 4.342827    & 2.3523445    & 2.830406    &  7.588481   \\
 & & & & \\
$P_B^a$ (d)  & 0.119(3)     & 0.41(2)     & 0.1322(16)   & 1.24(14)    &  1.12(3)\\
$P_B^b$ (d)  & 0.121        & 0.43        & 0.1330       & 1.201724(3) &
1.13 \\
 & & & & \\
$x^a$ (s)    & 0.0396(13)   & 0.49(4)     & 0.240(3)     & 0.79(14)    &  1.33(4)\\
$x^b$ (s)    & 0.0404       & 0.53        & 0.243        & 0.7659(14)  &  1.34    \\
 & & & & \\
$T_0$        & 51000.7979   & 51003.0587  & 51214.9496   & 51000.9649  & 51000.3173 \\
 & & & & \\
$N$          & 31 & 13 & 9 & 6 & 9 \\
\hline
\end{tabular}
\caption{ Comparison of the orbital parameters for several binaries in
the globular cluster 47~Tuc. The parameters indicated by an ``a'' were
determined by the method presented in this paper, the parameters
indicated by a ``b'' were determined by the timing analysis. The
latter are far more accurate (Freire et al. 2000) and their true value
is within half of the last significant digit quoted. For 47~Tuc~S, the
``b'' parameters are still derived from the period analysis.
The method presented in this paper is
accurate within the errorbars derived from Monte Carlo simulations
(Appendix A). $N$ is the number of individual
measurements of acceleration and period; for 47~Tuc~J, U and W these
were randomly selected from the many measurements of these pulsars
made by the 17.4-minute search; for 47~Tuc~S and T the data used
is in Table \ref{tab:detections}.}

\label{tab:parameters}
\end{center}
\end{table*}

With these new, improved $P_B$ and $T_{\rm asc}$, the
orbital parameters can be refined even further, using conventional
fitting to period and, after that stage, more precise periods derived
from Times-of Arrival of pulses (see Freire et al. 2000, section 2).
In section 3 of that paper we present a phase-coherent timing solution
for 47~Tuc~T; for 47~Tuc~S the best solution is:

$P$ = 2.830406030(3) ms,

$P_B$ = 1.20172465(7) days,

$x$ = 0.766338(9) s,

$T_{\rm asc}$ = 51000.964588(8) (MJD).

No eccentricity has been detected.


\section{Conclusion}
The procedure outlined in section \ref{sec:general} and
Appendix A is useful even when the binary's orbit
is well-sampled, because it provides automatically a ``first guess'' 
of the orbital parameters, which can then be used in conventional
period analysis. The quality of this guess obviously depends on phase
coverage and the precision of the measurement of accelerations. The procedure 
presented in this paper is, however, the only one capable of giving 
any estimates of the orbital parameters when the data samples the orbital 
period sparsely. For this purpose, the epochs of the individual 
period/acceleration measurements are completely irrelevant as long as the
measurements have reasonable orbital phase coverage; their values are
only needed to make the appropriate period corrections to the
barycentre of the solar system.

\section*{Acknowledgements}

We thank the skilled and dedicated telescope staff at Parkes for their
support during this project, and Vicky Kaspi, Froney Crawford, Ingrid
Stairs, Fernando Camilo and Jon Bell for assistance with observations
We also thank N. Wex for very stimulating discussions, from which
the techniques in Appendix A emerged. The Australia
Telescope is funded by the Commonwealth of Australia for operation as a
National Facility managed by CSIRO.  PCF gratefully acknowledges
support from Funda\c{c}\~{a}o para a Ci\^{e}ncia e a Tecnologia through
a Praxis~XXI fellowship under contract no.  BD/11446/97.



\appendix
\label{sec:appendix}
\noindent{\bf APPENDIX A}

Fitting an ellipse to a set of points can be quite
cumbersome, at least from a computational point of view. However,
eqns.~(\ref{eq:Pcirc}) and (\ref{eq:Acirc}) can be combined to derive a
simple expression, which is linear in the unknown parameters and hence
easy to fit. By squaring each of the equations, multiplying one of the
squares by an appropriate constant and summing them, we obtain the
simple equation of a parabola, given in the form
\begin{equation}
A^2 = a_2 P^2 + a_1 P + a_0,
\label{eq:parabola}
\end{equation}
where $A$ and $P$ represent acceleration and period. It is
straightforward to model the set of squares of the observed accelerations
and the corresponding periods, ($A_k^2$, $P_k$) by applying a linear
least squares fitting algorithm. We use the Singular Value Decomposition
method, as for instance described by Press et al.~(1992\nocite{ptvf92}), to
determine the coefficients $a_0$, $a_1$ and $a_2$. \footnote{We note as
a technical detail, that instead of using the original periods, $P_k$,
we use in fact a scaled version of those during the fit to avoid
numerical problems, i.e.~we use an expression $C\cdot(P_k-\bar{P})$, 
where $C$ is a  constant and $\bar{P}$ the mean of all measured $P_k$'s.}
The period and the orbital parameters are then given by

\begin{equation}
P_0 = - \frac{a_1}{2 a_2}
\label{eq:P0}
\end{equation}

\begin{equation}
P_B = \frac{2 \pi c}{P_0 \sqrt {-a_2}}
\label{eq:PB2}
\end{equation}

\begin{equation}
x = \frac{P_B}{2 \pi P_0} \sqrt{ P_0^2 - \frac {a_0}{a_2} }
\label{eq:X2}
\end{equation}

Despite the simplicity of the problem, there are large
covariances among the fitted parameters $a_0$, $a_1$ and $a_2$,
making it difficult to obtain reliable error estimates for the derived
parameters. For this reason, we employed a Monte-Carlo method in our
fitting algorithm, which provides not only precise values of the
orbital parameters but also reliable and realistic estimates of their
uncertainties. To achieve this, we generate a total of
10000 synthetic data sets, each
having the same number of observations as the original set of real
observations. For a given synthetic data set, the generated random
periods and accelerations are Gaussian distributed while centred on
the real observed data points, with dispersions given by the
observational uncertainties.

\begin{figure*}
\setlength{\unitlength}{1in}
\begin{picture}(0,2.5)
\put(-3.5,2.7){\includegraphics{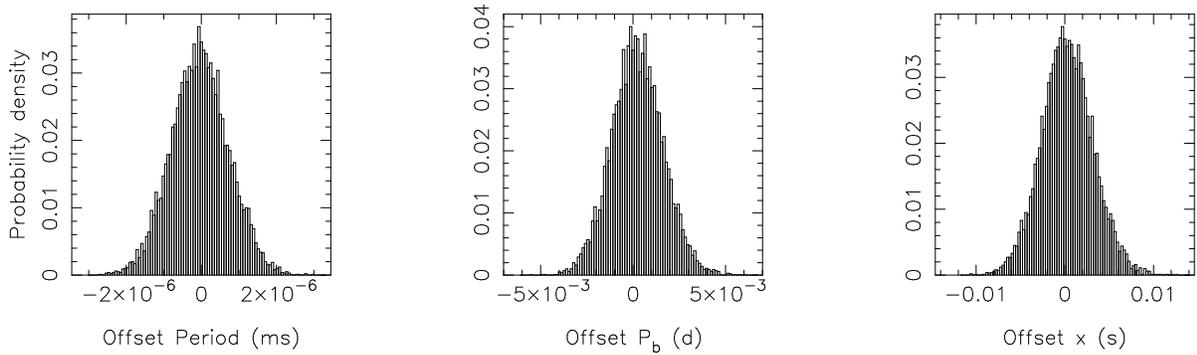}}
\end{picture}
\caption
[]
{Distributions of the parameters obtained by fitting 10000 synthetic
data sets generated in the Monte Carlo simulation in the case of 47~Tuc~W.
The values for the orbital parameters so determined are accurate to
within 1\%. Each distribution shows the offset from the best fit value.}
\label{fig:distributions}
\end{figure*}

We perform the described fit algorithm for each synthetic data set
and obtain probability distributions for the orbital parameters as shown in
Figure \ref{fig:distributions} for the case of 47~Tuc~W.  We choose
the maximum of each probability distribution as our best fit value of
a given orbital parameter and calculate its uncertainty from the
spread in its distribution.  The reliability of this method is
demonstrated in Table \ref{tab:parameters} where we compare the
parameters obtained using this method (columns 2 to 4) with the
parameters as determined in the subsequent timing analysis (columns 5
to 8). The timing
method provides far more precise values, however the comparison
clearly demonstrates that our new method provides accurate values
and calculates reasonable error estimates.

\end{document}